\documentstyle[aps,prc]{revtex}
\begin{document}
\title
{Octupole deformations in actinides at high spins 
within the cranking Skyrme-Hartree-Fock approach} 
\author
{A. Tsvetkov$^1$, J. Kvasil$^1$ and R.G. Nazmitdinov$^2$}
\address
{\it $^1$Particle and Nuclear Physics Institute, 
Charles University\\
V Hole\v sovi\v ck\'ach 2, 180 00 Praha 8, Czech Republic\\
$^2$Bogoliubov Laboratory of Theoretical Physics, 
Joint Institute for Nuclear Research\\
141980 Dubna, Moscow Region, Russia}
\maketitle

\begin{abstract}
The cranked Skyrme III effective Hamiltonian is applied for the analysis
of the rotational dependence of the quadrupole and octupole moments
in Ra, Th, and U isotopes. 
A comparison of the intrinsic electric dipole moments calculated
in the model with available experimental and theoretical values   
is presented.
It is found that the non-axial octupole deformation $Y_{32}$
becomes favorable at high spins for the actinide nuclei.
\end{abstract}

\vskip 0.3 cm

PACS number(s): 21.60-n; 21.60.Jz; 21.10.Ky; 27.90.+b

\vskip 0.5cm

In the last decade extensive experimental data have been obtained on
low-lying states of negative parity (see for review \cite{Ah,But96}). 
For the $Ra-Th (Z \sim 88$, $N\sim 134$)
and $Ba-Sm (Z\sim 56$, $N \sim 88$) nuclei low $3^-$ states, parity doublets, 
alternating parity bands with enhanced dipole transitions have been found. 
Since the octupole interactions are strongest when pairs of orbitals
from  the intruder sub-shell $(l,j)$ and the normal parity sub-shell 
$(l-3,j-3)$ are near the Fermi surface, the octupole correlations are
expected to become important for these nuclei \cite{BM,Ab}.

The features observed in nuclei are very similar to ones familiar 
from molecular physics.
In molecules a stable octupole deformation 
leads to the appearance of the rotational bands with the alternating
parity levels connected by strong E1 intra-band transitions.
M\"oller and Nilsson
used the assumption about a symmetry breaking 
of the intrinsic reflection symmetry of the 
single-particle potential
and demonstrated 
the instability with respect to the axial octupole
deformation in the Ra-Th region \cite{Mol70}.
The analysis was performed within the macroscopic-microscopic 
approach that combines 
the liquid drop model and the Strutinsky shell correction method 
for a modified harmonic oscillator potential.
Later, within this approach numerous results, 
based on various single-particle 
phenomelogical potentials, have been reported regarding 
the presence of a stable axial octupole deformation in the ground state 
of some nuclei in the actinide region \cite{But96}. 
The macroscopic-microscopic approach was also used in 
the combination with the cranking model to analyze
properties of rotating nuclei with a stable
axial octupole deformation  (see e.g. \cite{Naz85,Fra84}).
The problem of non-axial octupole deformations, being
for a long time a {\it terra incognita}, 
attracts more attention during last years.
The study of the non-axial octupole deformations 
could shed light on the tendency for the system
on the way toward fission to avoid the superdeformation \cite{BM}. 
Calculations by the macroscopic-microscopic method with the
Woods--Saxon potential predicted the importance of the
banana-type $Y_{31}$ deformation for highly deformed nuclei
\cite{Ch}, while the $Y_{32}$ deformation was found to be
important in the $^{222}Ra$ nucleus \cite{Dud}.
Manifestations of pronounced shell effects have been discovered  
when non-axial octupole deformations are added to a harmonic 
oscillator model \cite{He98}.

Although the macroscopic-microscopic approach is a very efficient practical 
tool for the analysis of shell effects in finite Fermi systems, 
basically, it is
a  phenomenological method which is unable, for example, 
to predict binding energies.  
Development of microscopic approaches, 
which start with the effective
nucleon-nucleon interactions treated within a 
self-consistent Hartree-Fock (HF) method, opened a new avenue 
in the nuclear structure studies \cite{Ab,But96}.
Using the Skyrme III effective interaction, Bonche et al. \cite{Bon86}
did the first self-consistent calculations for the ground states of 
the Ra-Th region. It was found that the total energy has a minimum 
for the nonzero value of the $Q_{30}$ momentum in $^{222}Ra$. 
Robledo and Egido \cite{Egi90} employed the Hartree-Fock-
Bogoliubov method with the Gogny interaction and found 
the axial octupole deformation in the ground states
in $^{222,224}Ra$, $^{222}Th$ and $^{142-148}Ba$.
Recent self-consistent HF+Skyrme calculations 
\cite{Tak98}  suggest that
the oblate states in nuclei with  $A\sim80$ are soft against the
$Y_{33}$ deformation in the ground states. The self-consistent
cranking calculations based upon the HF+Skyrme \cite{Yam98} and the 
HF+Gogny \cite{Tak} approaches 
predict that the $Y_{31}$ deformation is important for a correct 
description of the yrast band of $^{32}S$.
However, the systematic investigation of the non-axial 
octupole deformations in the actinide region at high spins is  
missing. On the other hand, 
the calculations in the framework of the $spdf$ IBM model show
that the $1-pf$ boson limit (dipole-octupole vibrations) can describe
essential features related to the octupole bands in
the Rn-Th region \cite{Zam}.
Using the quasiparticle spectrum generated by the mean field with the
reflection symmetry,   
the quasiparticle-phonon model reproduced quite well the
experimental data for $B(E1)$-transition probabilities in
$^{227-231}$Ra and $^{229-231}$Th \cite{KJ}.
As is seen from this discussion, the question of the existence of 
stable octupole deformations in the actinide region 
at different angular momenta is still controversial and 
challenging for microscopic approaches. 
In this paper the dipole, quadrupole and 
octupole intrinsic moments of rotating nuclei from
the Ra-Th-U region
are studied within the HF+Skyrme approach.
To this aim we exploit the novel version (1.75) of the code 
HFODD \cite{Dob97} for the Skyrme III force \cite{Bei75} 
in the particle-hole channel. 
The code was successfully tested for the analysis of 
superdeformed nuclei (see, for example, \cite{Dob95}).
Therefore, the basis employed by the code is sufficient
for our study.              
In the present calculations the pairing correlations
are neglected. They are not expected to play an important role at 
large rotational frequencies, however, they may influence 
results obtained for the ground state. 

We briefly remind the main ingredients of 
the calculations performed with the code HFODD, since it will be useful 
for our discussion below.
The code \cite{Dob00} solves the Hartree-Fock problem
in rotating frame with the cranking Hamiltonian in the form
\begin{equation}
H_{\Omega}=T+V+V_{coul}-\Omega \hat I_{y} - \sum_{\tau=n,p}
\lambda_{\tau} \hat N_{\tau} + \sum_{\lambda \mu} C_{\lambda \mu}
(\hat Q_{\lambda \mu} - \bar Q_{\lambda \mu})^{2}.
\label{(4)}
\end{equation}
Here $T$ represents the kinetic energy operator, $V$ is the Skyrme
potential, $V_{coul}$ stands for the Coulomb term. The quantity
$\Omega$ is the rotational frequency, $\hat I_{y}$ is the total angular
momentum projection onto the cranking axis (y-axis); $\lambda_{\tau}$ 
and $\hat N_{\tau}$ are the chemical potential and the particle number 
operator, respectively, for protons ($\tau=p$) and neutrons ($\tau=n$). 
The last term in (\ref{(4)}) 
represents  multipole constraints. It was introduced in order to find
a good minimum of the total energy that corresponds 
to the expected symmetry of the mean field (see below). 
$\hat Q_{\lambda \mu}$ is
the mass-multipole-moment operator (with the multipolarity $\lambda$
and the projection $\mu$), $\bar Q_{\lambda \mu}$ is the expectation value
of the corresponding multipole moment for a given rotational frequency 
$\Omega$. The stiffness parameter $C_{\lambda \mu}$ determines the weight
(or our priority) of the constraint for a given $\lambda$
and $\mu$ during the minimization process. The detailed 
description of each term in (\ref{(4)}) can be found in \cite{Dob97,Dob00}.
Minimizing the expectation value of the cranking 
Hamiltonian $H_{\Omega}$, 
$<HF;\Omega \mid H_{\Omega} \mid HF;\Omega>$ , for a 
given rotational frequency $\Omega$ and a
fixed set of the Skyrme parameters, the yrast state and equilibrium
values of mass-multipoles momenta, $<HF;\Omega \mid \hat Q_{\lambda \mu}
\mid HF;\Omega>$ can be found.
The knowledge of possible symmetries of a searched state could simplify
the minimization procedure. We used the following options of the code:
\begin{description}
\item[i.]  The mean field is supposed to be symmetric with
respect to the transformations $\hat S_{y}=\hat P e^{-i\pi \hat I_{y}}$,
$\hat S^{T}_{x}=\hat T \hat P e^{-i\pi \hat I_{x}}$, $\hat S^{T}_{z}=
\hat T \hat P e^{-i\pi \hat I_{z}}$ ($\hat T$ is the time reversal operator,
$\hat P$ is the parity operator). Such a situation corresponds 
to the standard $R$-invariance \cite{BM} when the expectation value, 
$<HF;\Omega \mid \hat Q_{\lambda \mu} \mid HF;\Omega>$, 
is nonzero only for even values 
of $\lambda$  and $\mu$. These symmetries are characteristic of
well-deformed nuclei without any octupole deformations.  
There are three symmetry planes ($y-z$,$x-z$,$x-y$) of the mean field.
\item[ii.] The only mean field symmetry is $\hat S_{y}=\hat P e^{-i\pi
\hat I_{y}}$ (there is only one $x-z$ symmetry plane). This is a general case
when all expectation values $<HF;\Omega \mid \hat Q_{\lambda \mu}
\mid HF;\Omega>$ can be nonzero. 
\end{description}
In the case {\bf i)} the quantum
numbers signature $r= \pm i$ (an eigenvalue of the operator
$\hat R_{y}= e^{-i\pi \hat I_{y}}$) and parity $\pi =\pm 1$ can be ascribed 
to each HF single-particle state.
In the case {\bf ii)} the parity is not a
good quantum number and the HF single-particle state can be 
characterized by the quantum number simplex:
$s=\pm i$ (an eigenvalue of the operator $\hat S_{y}$) 
(see a discussion about different symmetries in \cite{Dob99}). 
In each cases discussed above one can reduce the number of iterations
in the HF problem by the appropriate
constraint value $\bar Q_{\lambda \mu}$ in the last term in (\ref{(4)}).
In addition, the last term in (\ref{(4)}) allows to
avoid physically meaningless "deformations" which could correspond,
for example, to a shift of the nucleus center of mass. 
In order to ensure the translation  
invariance of the Hamiltonian 
we used the constraints with $\bar Q_{10} = \bar Q_{11} = 0$ and
the constraint $\bar Q_{21} = \bar Q_{2-1}= 0$ for the 
principal-axis condition.

The calculations for three chains of isotopes (Ra, Th, and U) were 
performed without imposing a restriction
on the axial reflection symmetry (the option {\bf ii)}).
For each nucleus two minima of the total energy
were obtained: one corresponds to nonzero octupole moments and the
second minimum is related to the mean field solution with 
the conserved reflection symmetry  (the option {\bf i)}). 
In order to obtain the
reflection asymmetric minimum the optimal values for 
the constraint $\bar Q_{30}=10^{2} fm^{3}$
and for the stiffness parameter $C_{30} \approx 0.01$ were found and 
used in the calculations. 
The constraints for the other octupole moments 
result in the smaller binding energies.
Both the  minima of the total 
energy are very close. For instance, for a well octupole deformed
nucleus $^{226}Ra$ the octupole deformed minimum is lower by
1 MeV with respect to the reflection-symmetric minimum. 
All plots presented 
on Figs.1-4 are related to the reflection asymmetric minima. 
From these figures it follows that 
the octupole deformed Ra, Th, and U isotopes
have axial symmetric shapes at  small rotational frequencies $\Omega$. 
With the increase  of 
the rotational frequency both the non-axial moments $Q_{22}$ and $Q_{32}$ 
increase whilst the axial moments $Q_{20}$ and $Q_{30}$ decrease
for all three elements studied. 
The exceptions are provided
by the lightest isotopes of Ra and Th ($^{218-220}Ra$, $^{218-220}Th$).
In these isotopes the moment $Q_{30}$  
reaches the maximum at $\Omega \sim 0.3 MeV$ and then decreases.
Similar tendency regarding the $Q_{30}$ moment 
has been predicted for  even-even
 rotating isotopes $^{220-228}$Th in the macroscopic-microscopic 
approach \cite{But91}. However, 
in the calculations \cite{But91}
only axially symmetric shapes 
of the Woods-Saxon potential have been considered.
At larger rotational frequencies, we obtained that 
heavier isotopes, especially,  of uranium, 
behave as non-axial octupole deformed systems.
In Fig.3  the intervals of the rotational frequencies $\Omega$
for which the reflection asymmetric minimum lies below the reflection
symmetric one are denoted by arrows.  
From Fig.4 it follows that the non-axial octupole moment $Q_{32}$ 
increases with the increase of the rotational frequency 
in $^{221,225,228}$Ra, $^{223,229,231}$Th and $^{226,231,233-238}$U isotopes.
This predominance of the non-axial octupole deformation in some nuclei 
may be  tested by measurements of the $E1$ transitions. 
A small displacement between the center-of-mass and the center of charge,  
caused by the non-axial deformation, 
may enlarge the $E1$ transitions within a parity doublet. 
In Fig. 5 the induced electric dipole moment $D_{0}$,
\begin{equation}
D_{0}=e \lbrack \frac{N}{A} Q_{10}(p) - \frac{Z}{A} Q_{10}(n) \rbrack,
\label{(5)}
\end{equation} 
is shown for different mass numbers A and 
different values of the rotational frequency.
In (\ref{(5)}) the quantities $Q_{10}(p)$
and $Q_{10}(n)$ are the proton and neutron contributions to 
the mass-dipole-moment $Q_{10}$, respectively. 
As is seen in Figs. 3  and 5, the value of $D_{0}$ is correlated 
with the axial octupole moment $Q_{30}$. 
The larger the value of $Q_{30}$ the larger is the electric
dipole moment $D_{0}$ induced by the octupole deformation. This tendency 
is consistent  with all known results concerning the $D_{0}$ 
(see \cite{But96} and the references there). 
The non-axial octupole deformation $Q_{32}$
influences the value of $D_{0}$ as well, however, its contribution is
less essential. In the chain of isotopes, from the lighter to the heavier 
one, we obtained the sign change of the quantity  $D_{0}$, which preserves
at different rotational frequencies.
This behavior is similar to the one predicted in \cite{Egi90} with
the Gogny forces for Ra and Th isotopes without rotation. 
A small value of $D_{0}$ preserves to high rotational frequency in
$^{222-224}$Ra, $^{226-228}$Th and $^{232,233}$U. 
Small changes in the quantity $D_{0}$  over the full frequency
range is obtained in $^{222-224}$Ra, $^{226-228}$Th and $^{230-233}$U.
Notice that for $^{222-224}$Ra our results  are consistent 
with the experimental observation of the behavior of $D_0$ \cite{Coc}.
The noticeable changes in the  $D_{0}$ due to the rotation 
are obtained in $^{228,231,233,234}$Ra, $^{220,230,231,233,234}$Th 
and $^{223-227}$U. 
In Fig. 6 the calculated values of $\mid D_{0} \mid$ 
are compared with the results of the other approaches and with 
the experimental values extracted from the $B(E1)$-transition 
probabilities \cite{But96}.
We remind that these experimental data provide  
only the absolute value $\mid D_{0}\mid$. 
The comparison is done for two values of 
the rotational frequency $\Omega = 0 MeV$
(low spin region) and $\Omega = 0.1 MeV$ (high spin region).
One can see from Fig. 6 that our results are in agreement with the other
calculations and that they reflect a qualitative behavior of the
experimental data for Ra and Th isotopes. 
It would be interesting to test our prediction for U isotopes
within other approaches and by experimental measurements.
In our analysis the quantity $D_{0}$ 
is nonzero and has a physical meaning only
in the intervals denoted by the arrows in Fig.3.
However, the correlations caused by
residual interactions can also contribute 
to the values of the total energy \cite{Al} and multipole moments \cite{KN}. 
Consequently, although the main features  should be preserved, 
the range of the intervals may be modified.
The work in this direction is in the progress.

In summary, we analyzed the mass-multipole-moments 
$Q_{2 \mu}$ ($\mu=0,2$), $Q_{3 \mu}$ ($\mu=0,1,2,3$)
and the intrinsic electric dipole moment $D_{0}$ in the Ra-Th-U region.
The analysis has been done within the self-consistent 
cranking+HF approach using the
Skyrme III effective interaction. It was found
that the octupole moment $Q_{30}$ decreases whilst 
the moment $Q_{32}$ increases with the increase of the rotational 
frequency $\Omega$. A similar behavior is found
for the quadrupole moments $Q_{20}$ and $Q_{22}$.
The quantity $D_{0}$ changes the sign at $\Omega = 0MeV$ at
$A \sim 224$ (for Ra),  $A \sim 227$ (for Th), 
and at $A \sim 232$ (for U). This tendency preserves over
the whole range of values of the rotational frequency 
$0.0\leq \Omega \leq 0.6$ MeV. 

\vspace{0.2cm}
\leftline{\bf Acknowledgments}
Authors would like to express their gratitude to Jacek Dobaczewski for
providing  the (1.75) version of the HFODD code and for valuable discussions
concerning the application of his code. 
The work was supported in part by a Czech Republic Grant No. 202/99/1718 
and  by the Russian Foundation for Basic Research under Grant 00-02-17194.

{\bf Figure Capture}
\begin{description}
\item[Fig.1] The quadrupole moment $Q_{20}$ as a function of the rotational
frequency.  The results are displayed: for Ra-isotopes at the top panel,
for Th-isotopes at the middle panel and 
for U-isotopes at the bottom panel.
\item[Fig.2] The quadrupole moment $Q_{22}$.
Similar to the Fig.1
\item[Fig.3] The octupole moment $Q_{30}$. Similar to the Fig.1.
 See also the text.
\item[Fig.4] The octupole moment $Q_{32}$.
Similar to the Fig.1
\item[Fig.5] The electric dipole moment $D_{0}$ induced by the octupole 
deformation as a function of the rotational frequency.
Similar to the Fig.1
\item[Fig.6] 
Experimental (with error bars) and calculated intrinsic dipole
moment $\mid D_{0} \mid$.
Experimental and calculated values are indicated by symbols denoted: 
''els'' and ``tls'' for measurements and our results for
low-spin states ($I\sim 0 \hbar$), respectively;
``ehs'' and ``ths'' for measurements and our results
for higher spin states ($I\sim 8 \hbar$), respectively.
Experimental values are taken from \cite{But96}. The
symbols ``BNls'' and ``BNhs'' correspond to the results from
\cite{But91} obtained at  $\Omega =0MeV$ and $\Omega = 0.1MeV$, respectively.
The calculated values denoted by ``ER'' are  taken from \cite{Egi90}. 
\end{description}

\end{document}